# On finite size effects, ensemble choice and force influence in dissipative particle dynamics simulations

Miguel Ángel Balderas Altamirano*, Elías Pérez* and Armando Gama Goicochea**[1]

*Instituto de Física, Universidad Autónoma de San Luis Potosí, Avenida Álvaro Obregón 64, San Luis Potosí 78000, San Luis Potosí, Mexico
**División de Ingeniería Química y Bioquímica, Tecnológico de Estudios Superiores de Ecatepec, Av. Tecnológico s/n, Ecatepec 55210, Estado de México, Mexico
agama@alumni.stanford.edu

**Abstract.** The influence of finite size effects, choice of statistical ensemble and contribution of the forces in numerical simulations using the dissipative particle dynamics (DPD) model are revisited here. Finite size effects in stress anisotropy, interfacial tension and dynamic viscosity are computed and found to be minimal with respect to other models. Additionally, the choice of ensemble is found to be of fundamental importance for the accurate calculation of properties such as the solvation pressure, especially for relatively small systems. Lastly, the contribution of the random, dissipative and conservative forces that make up the DPD model in the prediction of properties of simple liquids such as the pressure is studied as well. Some tricks of the trade are provided, which may be useful for those carrying out high – performance numerical simulations using the DPD model.

---

[1] Corresponding author.

# 1 Introduction

Molecular Dynamics (MD) simulation is now a popular multidisciplinary research tool in science, which consists essentially of solving Newton's second law of motion for fluids made up of many particles using computers in discreet time steps to calculate properties of interest. Some systems can be easily modeled as a pure homogeneous fluid, but some others require the modeling of complex interactions competing with one another, as is the case in biological systems with many molecules interacting in solution such as proteins, viruses or molecular chaperones [1 – 3]. Computer evolution in recent years has already brought many opportunities for the modeling not only of toy models but also of useful realistic problems. It must be borne in mind though that all simulations are necessarily subject to a number of choices, such as the size of the cell and the time step, model interactions, and thermodynamic conditions under which the simulations are performed. Therefore, some guidelines must be followed to make those choices judiciously [1].

Choosing a box size depends on finding a compromise between the smallest size that requires the least time to complete the calculations, and the largest number of particles one can calculate to capture realistic behavior found in experiments. Another issue is time; some proteins need at least a few seconds to unfold; nowadays MD can simulate 100 ns with little effort, but much work is still required to go beyond that. To tackle these problems of scale, some techniques have been proposed that can increase the size or the observation time of the simulations. One of those techniques is known as Dissipative Particle Dynamics (DPD) [2], which is very similar in its essence to MD, namely solving the forces acting in many particle systems to obtain their momenta and positions. The difference between them comes from the choice of the model forces and of the thermostat. While MD requires usually of conservative forces only, DPD adds a dissipation force and a randomly fluctuating force to its structure, given respectively by eqs. (1) and (2) below:

$$F_{ij}^D = -\gamma \omega^D(rij)[\hat{\boldsymbol{r}}_{ij} \cdot \boldsymbol{v}_{ij}]\hat{\boldsymbol{r}}_{ij} \tag{1}$$

$$F_{ij}^R = \sigma \omega^R(rij)\varepsilon_{ij}\hat{\boldsymbol{r}}_{ij} \tag{2}$$

where $\boldsymbol{r}_{ij} = \boldsymbol{r}_i - \boldsymbol{r}_j$, $r_{ij} = |\boldsymbol{r}_{ij}|$, $\widehat{\boldsymbol{r}_{ij}} = \boldsymbol{r}_{ij}/r_{ij}$ ; $r_{ij}$ is the relative position vector between particles $i$ and $j$, $\sigma$ is the noise amplitude, $\gamma$ is the viscous force amplitude and $\boldsymbol{v}_{ij} = \boldsymbol{v}_i - \boldsymbol{v}_j$ is the relative velocity between the particles, with $\varepsilon_{ij} = \varepsilon_{ji}$ being random numbers with Gaussian distribution between 0 and 1, and unit variance. The weight functions $\omega^D$ and $\omega^R$ are related as follows:

$$\omega^D(r_{ij}) = [\omega^R(r_{ij})]^2. \tag{3}$$

The spatial dependence of the weight function can be freely chosen, as long as eq. (3) is fulfilled; for computational convenience only, $\omega^R = \left(1 - \frac{r_{ij}}{r_C}\right)$. The constants in eqs. (1) and (2) must obey the fluctuation – dissipation theorem, which yields [3]:

$$k_B T = \frac{\sigma^2}{2\gamma}, \qquad (4)$$

$k_B$ is Boltzmann's constant and $T$ the absolute temperature. When one chooses a value for those two constants the temperature is immediately set, thereby defining a built – in thermostat.

MD simulations usually resort to using the Lennard – Jones potential to model van der Waals – type of forces, complemented with the Coulomb equation for electrostatics to account correctly for the interactions of atoms and molecules. DPD on the other hand incorporates a repulsive ( $F_{ij}^C$ ) conservative, linearly decaying force, see eq. (5) below, which comes with some advantages: being a short range force, it makes its calculation between pairs of particles more efficiently performed by computers because the force changes little with distance and dies off beyond a cutoff distance. Secondly, the particles are represented by beads without internal structure, where the strength of the force between beads is obtained from the grouping of several atoms or even molecules of the fluid, through a coarse – graining procedure based on the chemical nature of the beads [4]. The conservative force is given by

$$F_{ij}^C = a_{ij}\left(1 - \frac{r_{ij}}{r_C}\right); \qquad (5)$$

$a_{ij}$ is the interaction parameter between DPD particles $i$ and $j$; $r_C$ is the cutoff radius, beyond which all interactions become equal to zero. It is because of the choice of force law given by eq. (5) that DPD is a mesoscopic simulation technique. There are, of course, capabilities that one may need for applications and which DPD does not possess; one of those is that the repulsive nature of the forces leads to the absence of the so – called van der Waals loop in the pressure – volume phase diagram, hence liquid – gas transitions cannot be modeled with DPD. Also, the presence of the dissipative force yields simulations that cannot be performed at constant energy, which is necessary for the simulation of heat transport. However, both of those disadvantages have been circumvented [5, 6], at the cost of decreasing the computational efficiency and simplicity of the method. For more details about the DPD methodology and applications see, for example, refs. [7, 8, 9, 10].

In this brief review we summarize recent results on the effects of the finite size of the simulation box on the prediction of properties such as the stress tensor anisotropy, interfacial tension, dynamic viscosity and disjoining pressure, under various statistical ensembles. Also we discuss the role of the dissipative, random and conservative forces in the calculation of the pressure of simple fluids, as well as their dependence on the size of the time step and of the size of the simulation box.

## 2 Finite size effects in equilibrium and dynamic properties

A very popular application of MD is for the prediction of interfacial and surface tension. To accomplish that one needs to calculate the components of the pressure tensor; if the interface between the fluids is perpendicular to the $z$ - axis the interfacial tension is obtained as:

$$\gamma = L_z \left[ \langle P_{zz} \rangle - \tfrac{1}{2} \left( \langle P_{xx} \rangle + \langle P_{yy} \rangle \right) \right], \tag{6}$$

where $\langle P_{xx} \rangle$, $\langle P_{yy} \rangle$ and $\langle P_{zz} \rangle$ are the diagonal components of the pressure tensor, averaged over time or over an ensemble. If one models a homogeneous liquid with periodic boundary conditions in all directions, where no interfaces are expected to appear, the quantity defined by eq. (6) can no longer be called an "interfacial" tension; it is only a measure of the stress anisotropy. To test the influence of finite size on the interfacial tension, Gama Goicochea and co − workers carried out simulations of a mixture of two immiscible model liquids as a function of the size of the simulation box using the standard DPD model, and using the Monte Carlo (MC) method under the $NP_{zz}T$ ensemble [11]. In this ensemble the component of the pressure tensor which is perpendicular to the interface separating the liquids ($P_{zz}$) remains constant, in addition to $N$ and $T$. The comparison between both methods is shown in Fig. 1; asterisks on quantities indicate they are expressed in reduced DPD units.

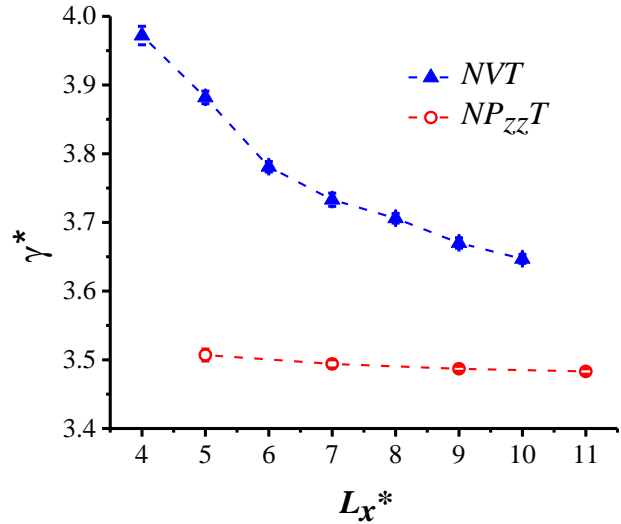

**Fig. 1** Interfacial tension results for a mixture of two DPD fluids, obtained from simulations carried out in the *NVT* (filled triangles) and *NP$_{zz}$T* (open circles) ensembles. The error bar sizes, shown in the figure, are of the order of the size of the symbols. Except for the choice of ensemble, both series of simulations have identical interaction parameters. The lines are only guides to the eye. All quantities are reported in reduced DPD units. Adapted from ref. [11].

As seen in Fig. 1, the choice of ensemble is very important when seeking to keep finite size effects to a minimum in the prediction of interfacial tension with DPD. Using standard DPD, that is, a dynamics that solves the equation of motion for the forces in eqs. (1), (2) and (5) in the canonical ensemble (*NVT*) leads to strong dependence of $\gamma^*$ on box size. This is to be contrasted with the choice of MC in the $NP_{zz}T$ ensemble, where $\gamma^*$ remains almost constant even for the smallest boxes. The difference between the $\gamma^*$ predictions at the smallest and largest boxes amounts to less than two percent, while the time required to carry out the simulations at those two sizes differ by about eighty percent! For this application it turns out that $NP_{zz}T$ is a better ensemble to simulate the interfacial tension than *NVT* because it is the one that more closely resembles the experimental conditions under which the interfacial tension is measured. Full details on this calculation of the interfacial tension between two simple liquids can be found in ref. [11], where the stress tensor anisotropy for a pure liquid with periodic boundary conditions was also calculated; it is shown in Fig. 2.

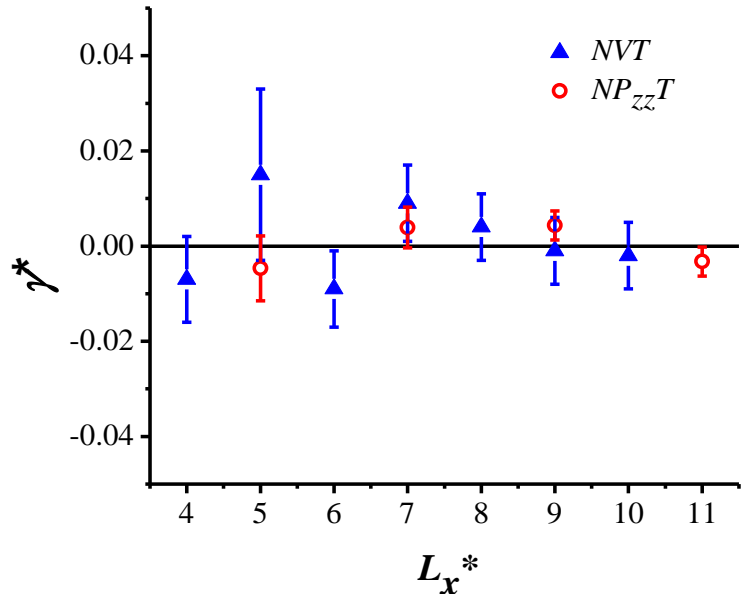

**Fig. 2** Stress anisotropy results for a pure, monomeric DPD model liquid in a simulation box with periodic boundary conditions using the *NVT* (filled triangles) and $NP_{zz}T$ (open circles) ensembles, as a function of box size. For a one phase system such as this $\gamma^*$ is expected to be zero (solid horizontal line) because there is no interface. Axes are in reduced DPD units. Adapted from ref. [11].

The stress anisotropy shown in Fig. 2 should be exactly equal to zero for a single liquid with no interfaces, as occurs for simulation boxes with periodic boundary conditions. However, numerical simulations as well as experiments are subject to finite size restrictions, which may introduce artifacts that shadow the value of the property in the thermodynamic limit. Although the values of $\gamma^*$ are small for both sets of data, those obtained with the $NP_{zz}T$ ensemble are closer to zero than those for the $NVT$ ensemble. Also, as the box's size is increased the stress anisotropy becomes closer to zero for both ensembles, meaning that finite size effects are negligible, as expected.

An attractive conservative force can be added to the original DPD interactions, leading to method that can predict liquid – vapor transitions [12]. Such term arises from the calculation of the local average density around each particle, giving rise to a conservative force that can be expressed as:

$$F_{ij}^C = a_{ij} w^C(r_{ij}) \hat{e}_{ij} + b_{ij}(\rho_i + \rho_j) w_\rho(r_{ij}) \hat{e}_{ij} \tag{7}$$

where the first term is the usual repulsive interaction, see eq. (5) and the second is the attractive contribution. The average local density around particle $i$ is $\bar{\rho}_i = \sum_{j \neq i} w_\rho(r_{ij}, R_d)$, $b_{ij}$ is the strength of the attractive force, and $w_\rho$ is a weight function given by

$$w_\rho(r_{ij}, R_d) = \frac{15\left(1 - r_{ij}/R_d\right)^2}{2\pi R_d^3} \tag{8}$$

where $R_d$ is the range of the weight function $w_\rho(r_{ij}, R_d)$. The addition of the second term in eq. (7) leads to a total conservative interaction that changes considerably with relative inter particle distance; this in turn produces oscillations in the pressure tensor components, as seen in Fig. 3.

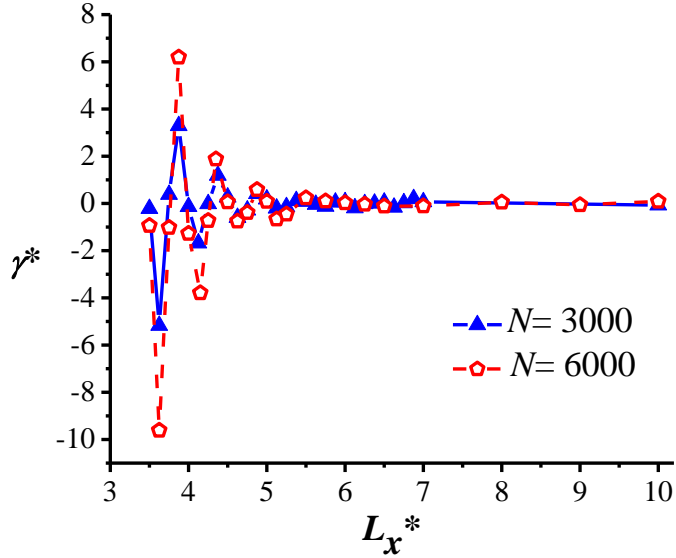

**Fig. 3** Stress anisotropy for a one component DPD fluid with attractive and repulsive interactions as a function of box side length, $L_x^*$. Results are shown for $N = 3000$ (filled triangles) and for $N=6000$ (open pentagons); for both the density is $\rho^*=3$. Lines are guides to the eyes. Both axes are represented in reduced DPD units. Adapted from [11].

The systems whose stress anisotropy are shown in Fig 3 are pure, monomeric liquids with periodic boundary conditions at two values of the particle number, hence $\gamma^*$ should be equal to zero. Yet one sees strong oscillations for both cases, at very small boxes, being larger for the largest number of particles (open pentagons in Fig.3). Two salient features in Fig. 3 demand comment: first, for boxes which are still relatively small, e. g. $L_x^*=5$, finite size effects in $\gamma^*$ are minimal and are once again negligible for moderately sized boxes. Therefore, even with the inclusion of an attractive term to the original conservative DPD term finite size effects are much smaller than those found in other interaction models. The origin of this feature can be traced back to the fact that the forces, even that in eq. (7) are short range. Secondly, one should note that the oscillations in Fig. 3, which come of course from oscillations in the components of the pressure tensor, have the same period for both systems. They originate from the basic interactions of the model and appear in all structural and thermodynamic properties of the fluid, such as in the radial distribution function and density profile, with the same period. This exponentially decaying oscillatory behavior is found in three – dimensional fluids interacting through short range potentials at high density, as predicted by the Fisher – Widom conjecture [12]. Full simulation details can be found in [11].

Let us now proceed to review the influence of finite size effects in non – equilibrium properties, particularly in the dynamic viscosity. To do so the following system was set up [14]: polymer chains were grafted at high density on parallel surfaces on the faces of the simulation cell perpendicular to the $z$ – axis; the solvent monomers were added explicitly also. To establish stationary Couette flow, a constant velocity was imprinted to the "heads" of the polymer chains grafted on the surface and an equal in magnitude but opposite in direction velocity was added to the grafted heads on the opposite surface. This setup forms a constant velocity gradient for the particles confined by the pore defined by the surfaces, along the $z$ – axis, which allows one to calculate the viscosity η as follows [15]:

$$\eta = \frac{\langle F_x(\dot{\gamma})\rangle/A}{\dot{\gamma}}. \tag{9}$$

In eq. (9) $\dot{\gamma}$ is the shear rate, which is given as $\dot{\gamma} = 2v/D$, where $v$ is the constant velocity applied to the grafted polymer heads, and $D$ is the distance between the surfaces, which is constant as well. The square area of each surface is $A$ and $\langle F_x(\dot{\gamma})\rangle$ is the mean force along the direction of the flow ($x$) that the particles on the surfaces experience, averaged over time for all particles. Figure 4 shows the dependence of the viscosity on the size of the simulation cell [14]. The difference between the value of the viscosity predicted for the smallest box and the largest one amounts to less than 0.5 percent only, showing once again that finite size effects are negligible in DPD even in the calculation of non – equilibrium properties. For additional discussion and computational details the reader is referred to [14].

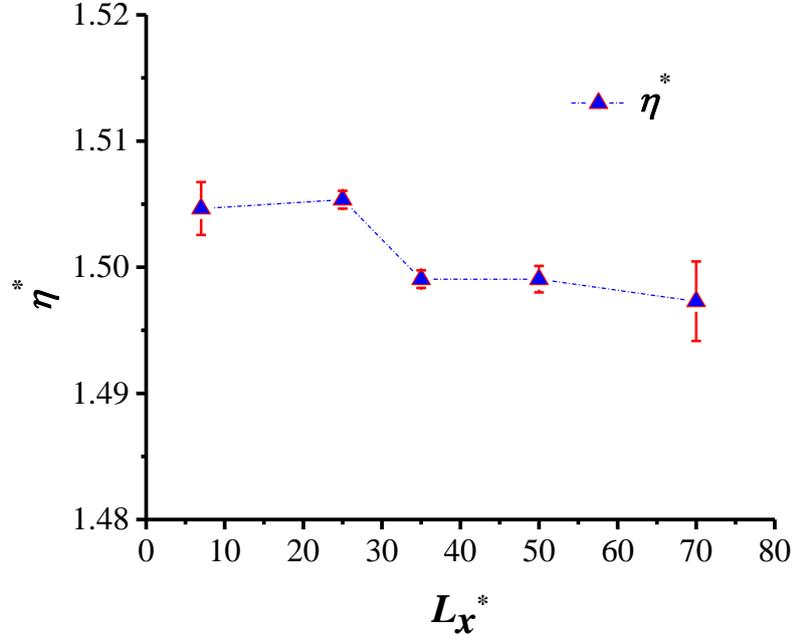

**Fig. 4** Finite size effect in the viscosity, $\eta^*$, of polymer brushes on parallel walls. The symbol $L_x^*$ represents the size of the simulation box in the $x$ direction, which is equal to that in the $y$ direction, $L_y^*$. In all cases, $L_z^* = 7$, the shear rate is $\dot{\gamma} = 0.28$, and the grafting density is $\Gamma^* = 0.30$. All quantities are reported in reduced DPD units. The line is only a guide to the eye. Adapted from [14].

## 3 Influence of the statistical ensemble choice in the prediction of pressure of confined fluids

Confined complex fluids are important for various reasons, among which is the need to understand how the internal structure of the confined fluid depends on its basic interactions and on the characteristics of the confinement. Additionally, from an industrially relevant point of view, many nanotechnology applications of confined fluids such as in the design of new stimuli – responsive material and in the fabrication of plastic sheets require detailed knowledge of properties such as the stability of fluid confined by surfaces. One key property of fluids under reduced symmetry conditions is the so called "disjoining" or solvation pressure ( $\Pi$ ) [16], defined as:

$$\Pi(z) = [P_{ZZ}(z) - P_B], \qquad (10)$$

where $P_{zz}$ is the component of the pressure tensor along the $z$ – axis, assuming that the confining walls are placed on the $xy$ – plane, and $P_B$ is the bulk pressure of the fluid, i.e., its pressure when it is not confined by the surfaces. If the fluid is not confined all diagonal components of the pressure tensor are equal to the bulk pressure and $\Pi = 0$. Therefore the disjoining pressure is a useful gauge of the stability of the confined fluid because is $\Pi > 0$ the walls are kept apart and stability ensues. If, however $\Pi < 0$ this signals attraction between the surfaces and the fluid collapses.

The calculation of equilibrium properties of fluids under confinement usually requires the implementation of the Grand Canonical (GC) ensemble, where the chemical potential must be kept constant, in addition to the volume and the temperature. This is necessary to ensure that the confined fluid is the chemical and thermal equilibrium with the virtual bulk fluid that surrounds the former fluid, and the mechanism used to reach equilibrium is through the exchange of particles between confined and bulk fluids. Implementing the GC ensemble requires performing averages over spatial configurations rather than averages over time, which in turn means one must carry out Monte Carlo simulations [17] instead of MD simulations. Implementing the DPD model interaction in the Grand Canonical Monte Carlo (GCMC) algorithm [18] allows one to test the influence of ensemble choice in the predicted value of $\Pi$ for DPD fluids.

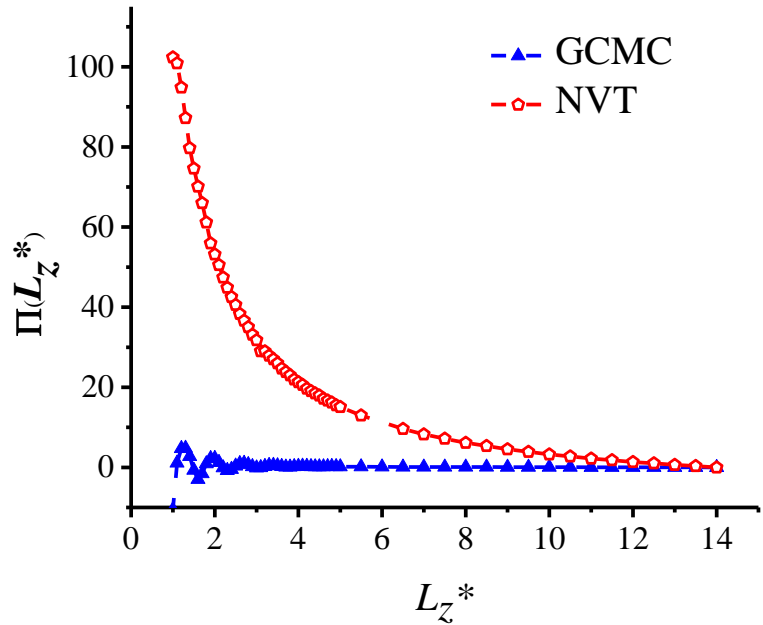

**Fig. 5** Comparison of the solvation or disjoining pressure of a pure monomeric confined fluid as a function of box size obtained at fixed chemical potential (GCMC, filled triangles), and a fixed density (NVT, empty pentagons). The interaction parameters, density and temperature are the same for both systems. Both axes are represented in reduced DPD units; lines are guides for the eye. Adapted from [19].

Figure 5 shows the comparison between the predictions of Π under the canonical ensemble (NVT) and under the GC ensemble for a simple monomeric fluid confined by structureless walls as a function of the simulation box size. The difference between those two approaches is striking at small to medium box sizes, becoming negligible only for the largest boxes. Moreover, the ergodic theorem states that the value of a property in equilibrium obtained from averages over time (as in NVT) must be the same as the value of it obtained from averages over configurations (as in GCMC) [20]. This is clearly not the case here, except for the largest boxes, as Fig. 5 shows. It should be stated at the outset that this is not a shortcoming of the DPD model, since similar results have been obtained for other models [21]. The reason for the discrepancy of predictions between those two ensembles lies in the fact that the NVT ensemble does not allow for the density fluctuations that the fluid undergoes when it is compressed, which bring also pressure fluctuations, while the GC ensemble does. That is why the pressure grows as the fluid is compressed under NVT conditions in Fig. 5, while fluctuations appear when pressure is calculated using the GCMC method [19]. Experiments on fluids under confinement using the surface force apparatus or atomic force microscopy confirm the predictions under the GC ensemble [22].

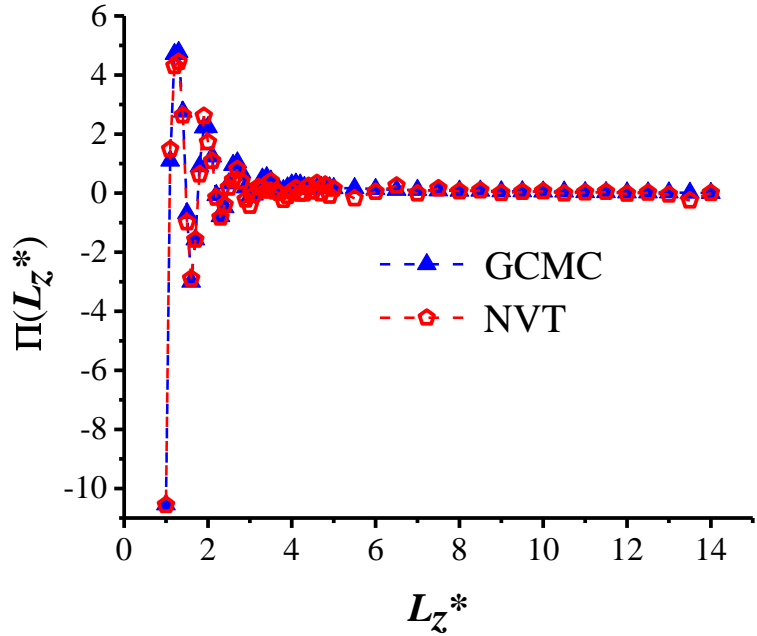

**Fig. 6** Comparison of the disjoining pressure of a confined monomeric liquid obtained at constant chemical potential (filled triangles; see GCMC curve in Fig. 5) with that obtained using standard DPD (NVT, empty pentagon) after having chosen the density of the latter to match the average density obtained at the same box volume from GCMC simulations. The scales on both axes are reported in reduced DPD units. Adapted from [19].

There is of course a caveat: by their very construction GCMC simulations are considerably more computationally intensive than MD simulations, regardless the interaction model [17], which in practical terms means that more computer time is required to predict properties in equilibrium with GC simulations. To try to get the best of both worlds, Balderas Altamirano and Gama Goicochea proposed a simple method to improve the speed of simulations of confined fluids without losing accuracy [19]. It consists of performing GCMC simulations at high confinement, keeping track of the average particle number; when that average is almost constant – which typically occurs after a few simulation blocks have been completed, the GCMC simulations are stopped. Then, that average particle number is inserted into the faster, NVT dynamics simulations, which are run until equilibrium is reached at a fraction of the computational cost [19]. The results, shown in Fig. 6, demonstrate that this simple method works very well, since the faster DPD simulations carried out under the canonical ensemble (NVT, empty pentagons in Fig. 6) reproduce the predictions of the disjoining pressure obtained from GCMC simulations (solid triangles in Fig. 6). The NVT data follow closely those from GCMC even for the smallest boxes, capturing the oscillations' amplitude and period as well. For full details and additional applications, see [19].

## 4   Influence of the conservative, dissipative and random DPD forces on the virial of simple fluids

In this section we review the influence that factors such as the size of the time step used in the integration of the equation of motion, simulation box size and strength of dissipation and Brownian motion have on the DPD forces' contribution to the virial calculation in simple liquids [23]. The focus is on the calculation of the virial only, because it is a popular method used to calculate the pressure in numerical simulations [24], as follows:

$$P = \rho k_B T + \frac{1}{3V} \langle \sum_{j>i} \vec{F}_{ij} \cdot \vec{r}_{ij} \rangle . \qquad (11)$$

In eq. (11) $P$ is the pressure of the homogeneous fluid, $\rho$ its density and the brackets represent average over time. The contribution of each DPD force to that term (second in eq. (11)) is calculated separately to study the influence of the above mentioned factors to the virial. Let us start with the influence of the time step while keeping the size of cell fixed; Fig. 7(a) shows the effect that increasing the time step has on the random force contribution to the virial, as a function of the simulation time, where $N$ is the number of times the equation of motion is solved numerically. Clearly, the random force contribution is very small, and quickly becomes negligible, independently of the choice of time step. This occurs because the random force is basically white noise. The dissipative force contributes very little to the virial also, as Fig. 7(b) shows, but in this case it requires longer simulation time, specially if the time step is very small, see the blue line in Fig. 7(b), where more than twenty thousand integrations of Newton's second law are needed for this artifact to become zero. It can be shown [23] that if the dissipative force is coupled to a random force, as occurs in the DPD model, the contribution of the former to the pressure – and to all equilibrium properties of the fluid – becomes zero, given a long enough period of time. The question is how long is long enough, but luckily if a relatively large time step is used, e.g. $\Delta t = 0.01$ the contribution of the dissipative force to the virial is zero almost immediately after the simulation has begun. When equilibrium is reached, only the conservative force should contribute to the virial, and this is indeed confirmed by Fig. 7(c).

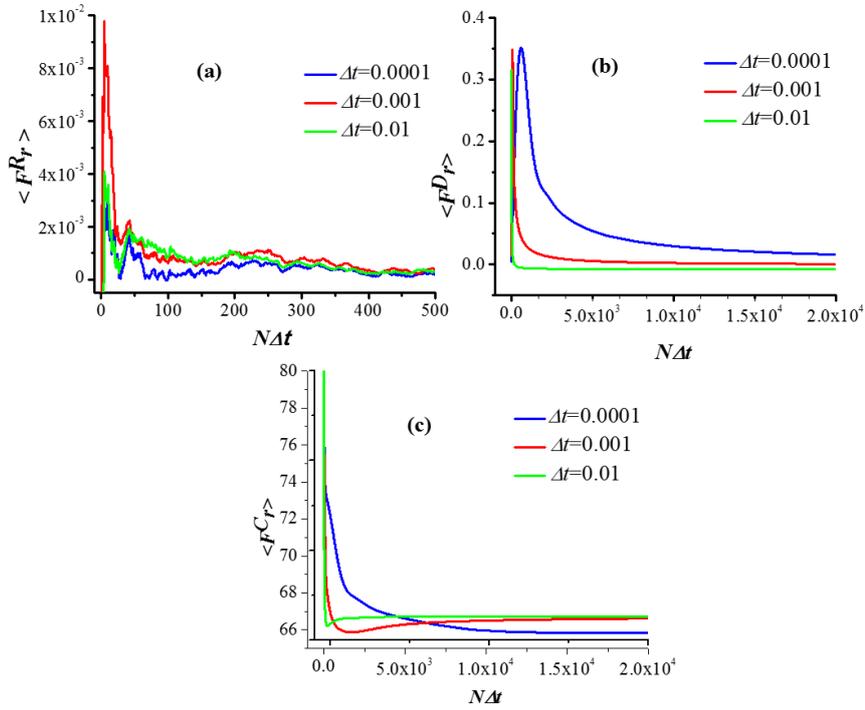

**Fig. 7** The virial contribution of (a) the random, (b) dissipative and (c) conservative forces for a DPD simple fluid as a function of time, for three choices of the integration time step. For the conservative force the interaction constant was chosen as $a_{ij} = 78.3$. The fluid is made up of 3000 identical particles in a cubic box with volume $V = 10 \times 10 \times 10$. All axes are shown in reduced DPD units. $N$ is the number of times the dynamics is solved. Adapted from [23].

Another variable of interest is the size of the simulation cell, whose influence on the virial is broken down into the three contributions shown in Fig. 8. In all cases the time step was set at $\Delta t = 0.01$, the box is cubic and the fluid's density is fixed at three, as in previous cases. Except for the small fluctuations that appear in the random force contribution to the virial in Fig. 8(a), what can be concluded from the results shown in Fig. 8 is that the size of the simulation box does not affect the contribution of the

DPD forces to the virial as the simulation time evolves. This is expected because of the short range nature of the DPD forces and because the virial contributions are traced as functions of time.

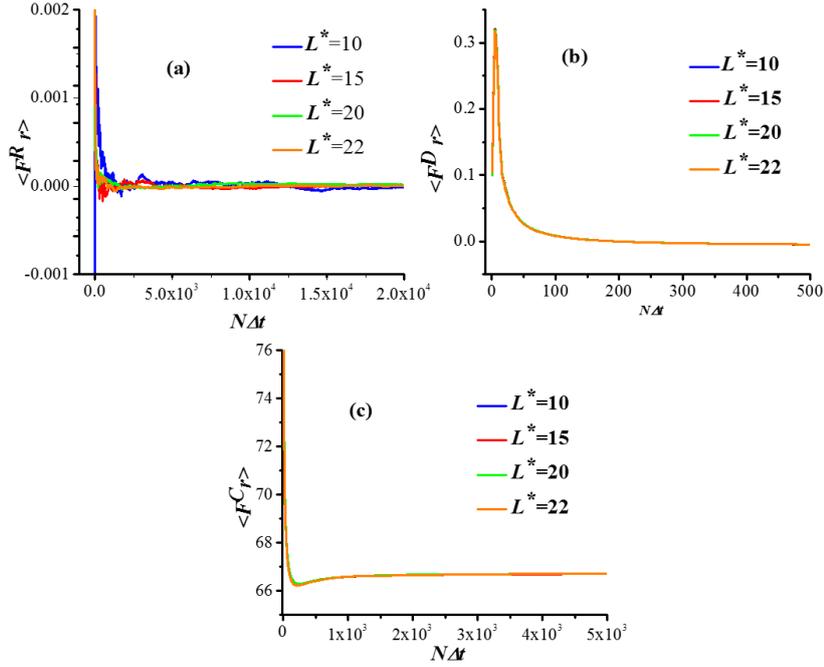

**Fig. 8** Effect of the size of the simulations box in the virial contribution of the (a) random, (b) dissipative and (c) conservative forces for a DPD fluid as a function of time, for four values of the volume of the cubic box with the side length $L^*$. The conservative force constant is $a_{ij}=$ 78.3. The fluid density is kept equal to $\rho^* = 3$ and the integration time step is chosen as $\Delta t = 0.01$ in all cases. All quantities are shown in reduced DPD units. Adapted from [23].

As a last case study we explore the dependence of the virial on the simulation time for four different values of the constants defining the strength of the dissipative and random forces in DPD, namely $\gamma$ and $\sigma$ in eqs. (1) and (2), respectively. For this part of the work the volume of the cell is fixed as well as the density and the time step in all simulations [23]. Increasing those constants amounts to making the fluid more viscous and the DPD thermostat comes into play to invest that increase into more Brownian motion.

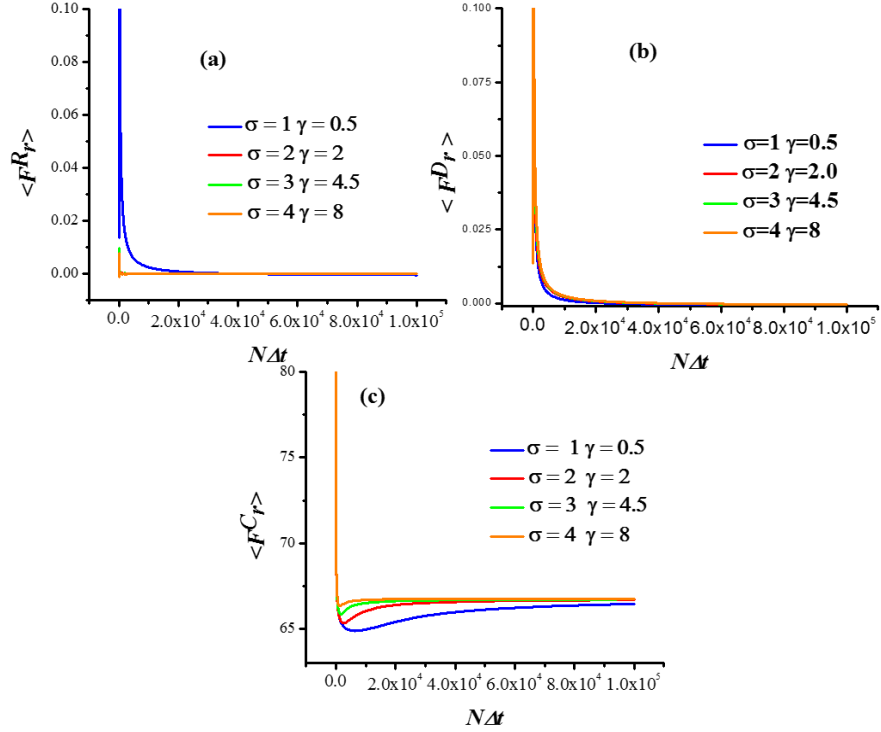

**Fig. 9** Effect of varying the strength of the random ($\sigma$) and dissipative ($\gamma$) forces in the contribution to the virial of the (a) random, (b) dissipative and (c) conservative forces for a monomeric DPD fluid as functions of time. In all cases the temperature is $T^* = 1$, the conservative force constant is $a_{ij} = 78.3$, and the fluid density is kept equal to $\rho^*=3$, while the integration time step was chosen as $\Delta t = 0.001$ in all cases. All quantities are reported in reduced units; adapted from [23].

As Fig. 9 shows, the more viscous the DPD fluid becomes, the less important the non − conservative interactions are with respect to their contributions to the virial as the numerical simulation evolves in time. Even for the least viscous fluid (blue line in Fig. 9), the dissipative and random forces' contributions to the virial are very small and the pressure can be accurately obtained from the conservative force contribution only. Moreover, the simulation running time required to remove artifacts contributing to the pressure from the dissipative and random forces is of the order of $3\times10^4$ except for the least viscous fluid, which is only of academic interest. As Fig. 9(c) shows, for long enough simulations only the conservative force matters in the calculation of the pressure through the virial, except for the least viscous fluid. The same trends are expected to hold for other thermodynamic and structural properties in equilibrium.

## 5     Conclusions

No technique is without shortcomings and DPD is no exception, but in this contribution we have focused on reporting some tricks of the trade that practitioners of numerical simulations might find useful, especially when it comes to trying to optimize computational efforts. By judiciously choosing parameters and scales, users of DPD can benefit from the fact that its mesoscopic reach and the simplicity of its interactions can be helpful tools to derive novel information on soft matter systems.

## Acknowledgments

MABA thanks PRODEP DSA/103.5/15/3894 and CA – Ingeniería de Procesos Químicos y Ambientales. MABA and AGG thank the Centro Nacional de Supercomputo (IPICYT) and the High Performance Computation Area of the Universidad de Sonora, for allocation of computer time; J Limón (IF UASLP) is acknowledged for technical support. AGG would like to thank JD Hernández Velázquez, J. Klapp, E. Mayoral and C. Pastorino for important discussions.